\documentclass[twocolumn]{revtex4}

\usepackage{ragged2e}
\usepackage{amsmath}
\usepackage{graphicx}
\usepackage{rotating}
\usepackage{bm}

\usepackage[usenames,dvipsnames]{color}

\definecolor{darkblue}{RGB}{0,0,196}

\begin{document}

\title{Random statistical analysis of transverse momentum spectra of
strange particles and dependence of related parameters on
centrality in high energy collisions at the LHC \vspace{0.5cm}}

\author{Xu-Hong~Zhang$^{1,}$\footnote{xhzhang618@163.com; zhang-xuhong@qq.com},
Fu-Hu~Liu$^{1,}$\footnote{Correspondence: fuhuliu@163.com;
fuhuliu@sxu.edu.cn},
Khusniddin~K.~Olimov$^{2,3,}$\footnote{khkolimov@gmail.com;
kh.olimov@uzsci.net},
Airton~Deppman$^{4,}$\footnote{deppman@if.usp.br}}

\affiliation{$^1$Institute of Theoretical Physics, State Key
Laboratory of Quantum Optics and Quantum Optics Devices \&
Collaborative Innovation Center of Extreme Optics, Shanxi
University, Taiyuan 030006, China
\\
$^2$Physical-Technical Institute of Uzbekistan Academy of
Sciences, Chingiz Aytmatov str. $2^b$, 100084 Tashkent, Uzbekistan
\\
$^3$National University of Science and Technology MISIS (NUST
MISIS), Almalyk branch, Almalyk, Uzbekistan
\\
$^4$Instituto de F{\'i}sica, Universidade de S{\~a}o Paulo, Rua do
Mat{\~a}o 1371-Butant{\~a}, S{\~a}o Paulo-SP, CEP 05580-090,
Brazil}

\begin{abstract}

\vspace{0.5cm}

\noindent {\bf Abstract:} We have studied the transverse momentum
($p_T$) spectra of the final-state strange particles, including
$K^{\pm}$, $\phi$, $\it\Xi$, and $\it\Omega$, produced in high
energy lead--lead (Pb--Pb), proton--lead ($p$--Pb), xenon--xenon
(Xe--Xe) collisions at the Large Hadron Collider (LHC). Taking
into account the contribution of multi-quark composition, whose
probability density distribution is described by the modified
Tsallis--Pareto-type function, we simulate the $p_T$ spectra of
the final-state strange particles by a Monte Carlo method, which
is shown to be in good agreement with the experimental data in
most the cases. The kinetic freeze-out parameters are obtained.
The present method provides a new tool for studying the spectra of
various particles produced in high energy collisions, reflecting
more realistically the collision process, which is of great
significance to study the formation and properties of the produced
particles.
\\
\\
\end{abstract}

\maketitle

\section{Introduction}

Plenty of final-state particles are produced in relativistic
heavy-ion collisions~\cite{1,2} in various collision processes,
resulting in different configurations of final-state particles. In
high energy collisions, novel physical phenomena can appear with
the most representative being formation of quark-gluon plasma
(QGP)~\cite{3,4}. In addition, multi-particle production provides
abundant information on thermal and statistical properties of a
system. Although quantum chromodynamics (QCD) can provide an
important theoretical basis to study strong interactions among
quarks and features of collision system~\cite{5,6}, thermal and
statistical method is still a powerful tool for such analysis.

Investigation of the transverse momentum ($p_T$) (or transverse
mass, $m_T$) spectra of the final-state particles is an effective
and fast method~\cite{1,2} to study the processes of multiparticle
production and system evolution. This method can help to obtain
the thermodynamic parameters of the final-state particles and
collision system. The $p_T$ spectra of different final-state
particles have been studied extensively. In addition to baryons,
leptons and other elementary particles have also attracted much
attention.

In recent years, the production of single- and multi-strange
particles~\cite{7,8,9,10} has attracted an increased interest,
being extensively researched. Based on the exotic
properties~\cite{15}, the $p_T$ spectra of strange particles
measured in high energy collisions by various collaborations have
been analyzed and predicted utilizing Tsallis
function~\cite{16,17,18,19,20,21}, non-equilibrium chemical or
kinetic freeze-out model~\cite{8}, etc.

Relative to pions, enhanced production of multi-strange
hadrons~\cite{10,22,23,24,25} in high-multiplicity proton--proton
collisions has been observed experimentally for the first time,
and it is found that the integrated yields of strange particles
increases significantly with the event charged-particle
multiplicity. Besides, the azimuthal angular correlation and the
mass dependence of $p_{T}$ due to the existence of QGP in
high-energy collisions have been also reported. Further researches
on strange particles can better explore and reveal the properties
of QGP.

We are interested in exploration of the properties of different
strange particles. In contrast to our previous
work~\cite{27,28,29}, in this work we adopt a new algorithm to
simulate and analyze the $p_T$ spectra of strange particles. We
aim to restore the collision process more realistically and
extract more accurate characteristic parameters. The multi-quark
composition of baryons is considered. A Monte Carlo
method~\cite{26,36,37} is used to simulate different transverse
momenta carried by various quarks. The modified
Tsallis--Pareto-type function~\cite{27,28,29} is used to define
the contribution of multi-quarks, from which the kinetic
freeze-out temperature ($T_{0}$)~\cite{30,31,32,33}, average
transverse flow velocity ($\langle\beta_{t}\rangle$)~\cite{34,
35}, and other related parameters can be directly extracted.

This paper is structured as follows. The formalism and Monte Carlo
method are briefly introduced in Section 2. The simulated or
fitted results, their comparison with the data, and discussion are
given in Section 3. Finally, in Section 4 we summarize our main
observations and conclusions.

\section{Formalism and method}

To deal with $p_{T}$ of strange particles produced in high energy
collisions, we briefly introduce the analysis, based on the Monte
Carlo method. Generally, the modified Tsallis--Pareto-type
function~\cite{27,28,29}
\begin{align}
f(p_T)=\frac{1}{N}\frac{dN}{dp_T}=C p_{T}^{a_{0}}
\bigg(1+\frac{m_{T}-m_{0}}{nT}\bigg)^{-n}
\end{align}
is suitable to characterize the $p_T$ spectra of final-state
particles in low- and intermediate-$p_{T}$ regions. Here, $N$ is
the number of particles, $C$ is the normalization constant, $T$ is
the effective temperature of a collision system, $n$ is an
entropy-related index, which is used to describe the degree of
non-equilibrium of the system, $m_{T} =\sqrt{p_{T}^{2}+m_{0}^{2}}$
is the transverse mass of a particle, $m_0$ is the rest mass of
the particle, and $a_{0}$ is the correction index, which makes the
function to fit better the spectra in low-$p_{T}$ region.

It should be noted that when we set $a_{0}=1$, Eq. (1) is
naturally converged to the Tsallis--Pareto-type function. That is
to say, the introduction of $a_0\neq1$ in Eq. (1) is a
modification of Tsallis--Pareto-type function~\cite{21}. To
extract characteristic parameters such as the thermal or kinetic
freeze-out temperature $T_{0}$ of a collision
system~\cite{30,31,32,33} and the average transverse flow velocity
$\langle\beta_{t}\rangle$ of the produced particles~\cite{34,35}
at the quark level, following Refs.~\cite{38,39,40,41}, we perform
the Lorentz-like transformation on $m_{T}$ and $p_{T}$ in Eq. (1)
by using
$m_{T}\rightarrow\langle\gamma_{t}\rangle(m_{T}-p_{T}\langle
\beta_{t}\rangle)$ and
$|p_{T}|\rightarrow\langle\gamma_{t}\rangle|p_{T}-m_{T}\langle\beta_{t}\rangle|$,
where $\langle\gamma_{t}\rangle=1/\sqrt{1-\langle\beta_{t}\rangle
^{2}}$ is the Lorentz factor.

Thus, the transverse momentum $p_{Ti}$ of the $i$-th quark
contributed to the particle $p_{T}$ is assumed to obey the new
modified probability density function
\begin{align}
\begin{split}
f_{i}(p_{Ti}) & = C_{i}
\frac{\langle\gamma_t\rangle^{a_0+1}}{m_{Ti}}
\big(m_{Ti}-p_{Ti}\langle\beta_t\rangle \big)
\big|p_{Ti}-m_{Ti} \langle\beta_{t}\rangle \big|^{a_0}\\
&\quad \times\bigg[1+\frac{\langle\gamma_{t}\rangle(m_{Ti}-p_{Ti}
\langle\beta_{t}\rangle)-m_{0i}}{nT_{0}}\bigg]^{-n}.
\end{split}
\end{align}
The temperature parameter at this time has been naturally
converted from $T$ in Eq. (1) to the kinetic freeze-out
temperature $T_{0}$. In Eq. (2), $m_{0i}$ is the constituent mass
of the $i$-th quark. As shown in the literature~\cite{42}, we have
$m_{01}=0.31$ GeV/$c^{2}$ for up and down quarks and $m_{02}=0.5$
GeV/$c^{2}$ for a strange quark. Thus, we have constructed the
probability density function, which satisfies multi-quark states.

Within the defined interval $[0,\infty)$ of $p_{Ti}$ of the $i$-th
quark, we have the normalization condition
\begin{align}
\begin{split}
\int^{\infty}_{0}f_{i}(p_{Ti})dp_{Ti}=1,
\end{split}
\end{align}
where the upper limit is usually quite a large value, but not
infinite. In a Monte Carlo calculation, $R_i$ denotes the random
number in $[0,1]$ and we have the relation satisfied by $p_{Ti}$
to be
\begin{align}
\begin{split}
\int^{p_{Ti}}_{0}f_{i}(p'_{Ti})dp'_{Ti}<R_{i}<
\int^{p_{Ti}+\delta_{p_{Ti}}}_{0}f_{i}(p'_{Ti})dp'_{Ti},
\end{split}
\end{align}
where $\delta_{p_{Ti}}$ is a small shift from $p_{Ti}$. According
to Eq. (4), one can obtain a series of discrete values of $p_{Ti}$
which satisfy Eq. (2). In the calculation, we take
$\delta_{p_{Ti}}=0.01$ GeV/$c$. Considering the contributions of
multi-quarks, we have $i=1$ to 2 for mesons and $i=1$ to 3 for
baryons, because meson consists of two quarks, and baryon has
three quarks.

In general condition, the particle transverse momentum $\bm p_T$
is the vector superposition of $\bm p_{Ti}$ of two or three
quarks. In the right-handed Cartesian coordinate system $O$-$xyz$,
let the $Oz$ axis be the beam direction, $xOz$ plane be the
reaction plane, and $xOy$ plane be the transverse one. In the
source rest frame, $\bm p_{Ti}$ is assumed to be isotropic. The
movement of the source along the $Oz$ axis constitutes the
longitudinal flow, and the interactions among different sources
causes the transverse flow. The longitudinal flow is not the focus
of the present work, though it can be described by the rapidity
distribution. The transverse flow is described by the average
transverse flow velocity $\langle\beta_{t}\rangle$ that is
mentioned in the above discussion.

For the $i$-th quark, we have the $x$- and $y$-components of $\bm
p_{Ti}$ to be
\begin{align}
\begin{split}
p_{Tix}&=p_{Ti}\cos\varphi_{i}=p_{Ti}\cos(2\pi r_{i}),\\
p_{Tiy}&=p_{Ti}\sin\varphi_{i}=p_{Ti}\sin(2\pi r_{i}),
\end{split}
\end{align}
where the isotropic azimuthal angle $\varphi_{i}$ is distributed
uniformly in $[0,2\pi]$. Using a random number $r_{i}$ distributed
uniformly in $[0,1]$, one has $\varphi_i=2\pi r_i$ in the Monte
Carlo calculation. According to the principle of vector
superposition, we obtain the expressions of $\bm p_T$'s components
and $p_T$ of the final-state particle to be
\begin{align}
\begin{split}
p_{Tx}&=\sum_{i=1}^{2,3}p_{Ti}\mathrm{cos}(2\pi r_{i}),\\
p_{Ty}&=\sum_{i=1}^{2,3}p_{Ti}\mathrm{sin}(2\pi r_{i}),\\
p_{T}&=\sqrt{p_{Tx}^{2}+p_{Ty}^{2}},
\end{split}
\end{align}
where the sums are considered due to the contributions of 2 or 3
quarks, the upper limit ``2" for the sums corresponds to the
meson, and the limit ``3" for the sums corresponds to the baryon
in the first two relations in Eq. (6).

According to the above analysis, we may obtain a lot of discrete
values of $p_{T}$ by the iterative calculations. Finally, we may
perform a statistical analysis on the discrete values of $p_T$ and
obtain suitable distributions such as $(1/2\pi p_T)d^2N/dydp_T$
and $d^2N/dydp_T$, which can be compared with the experimental
data, where $y$ denotes the rapidity and $dy$ is the width of
rapidity bin at mid-rapidity. In calculations, the statistical
interval of $p_T$ is taken to be 0.1 GeV/$c$.

Although we may simply use the modified Tsallis--Pareto-type
function to fit the experimental data and adopt the
$\chi^2$-minimisation scheme to obtain the fit parameters, the
results obtained by the simple method are at the particle level,
which seem not to be deeper insight compared to the results at the
quark and gluon level. To obtain the results at the
partonic-level, we may consider the contribution of multiple
partons to particle's transverse momentum. However, the analytical
expression obtained at the partonic-level is not available due to
the complex calculations for the superposition of multiple
transverse momenta with randomized azimuthal angles. Instead, we
may use the Monte Carlo method to obtain numerical results. This
is suitable for the case discussed in this paper.

It should be noted that several restrictions are used in the Monte
Carlo calculations. {\it Firstly}, the constituent masses of the
considered quarks are determinate. These masses are taken from the
literature~\cite{42}. {\it Secondly}, $p_{Ti}$ of a quark is
restricted to obey a given function. This function is modified
from the current Tsallis-like function. {\it Thirdly}, an
isotropic emission in the rest frame of emission source is
assumed. The momentum in the rest frame of emission source is then
obtained. {\it Fourthly}, the movement of emission source is
restricted by its rapidity which is assumed to be evenly
distributed in the projectile or target thermalized region in the
rapidity space. The rapidity of emission source is also related to
longitudinal flow, which restricts the rapidity range. {\it
Fifthly}, the conservation of energy and momentum is obeyed. The
combination of $p_{Ti}$ of 2- or 3-quarks with randomized
azimuthal angles into $p_T$ of hadron can be conducted, in which
any energy beyond hadron's rest mass is converted into hadron's
kinetic energy. {\it Sixthly}, the distribution of $p_T$ is
restricted by the experimental data.

\section{Results and discussion}

\subsection{Comparison with data}

In the study, we analyze the $p_T$ spectra of strange particles
generated in high energy lead--lead (Pb--Pb), proton--lead
($p$--Pb) and xenon--xenon (Xe--Xe) collisions at the Large Hadron
Collider (LHC). Figure 1 shows the $p_T$ spectra of $K^{+}+K^{-}$
(a), $(K^{*0}+\overline{K}^{*0})/2$ (b), $K_S^0$ (c), $\phi$ (d),
$\it\Lambda$ (e), $\it\Xi^-$ (f), $\it\overline{\Xi}^+$ (g),
$\it\Omega^-$ (h), and $\it\overline{\Omega}^+$ (i) produced in
Pb--Pb collisions with different centrality intervals marked in
the panels at the collision energy $\sqrt{s_{\rm NN}}=2.76$ TeV
per nucleon pair. The symbols represent the data measured by the
ALICE Collaboration~\cite{43,44,45}. The mid-(pseudo)rapidity
range of panel (a) is $|\eta|<0.2$, and those of other panels are
$|y|<0.5$. For better distinguishing and comparing, the data with
different centralities in different panels are multiplied by 10 to
the $n$-th power. In addition, the gray rectangles represent the
uncertainty of the experimental data, whose length is the
uncertainty of the abscissa, whereas the width is the uncertainty
of the ordinate, which denote the quadratic sum of the statistical
and systematical uncertainties. The solid curves represent the
Monte Carlo results calculated from the probability density
function in Eq. (2) and the $p_T$ expression in Eq. (6).

We use $\chi^{2}=\Sigma_{j}[(\mathrm{Data}_j-\mathrm{Fit}_j)^2/
\mathrm{Uncertainty}_j^2]$ to quantify the deviation of the Monte
Carlo results from the experimental data, where $j$ denotes the
order number of the data. The smaller the mean of $\chi^{2}$ is,
the closer the Monte Carlo results to the experimental data are.
The free parameters, the entropy-related index $n$, the kinetic
freeze-out temperature $T_{0}$, the correction index $a_{0}$, and
the average transverse flow velocity $\langle\beta_{t}\rangle$ are
extracted by the method of least squares. The values of free
parameters, $\chi^{2}$ and the number of degree of freedom (ndof)
are listed in Table 1. One can see that in most cases the Monte
Carlo results agree well or approximately with the experimental
$p_T$ spectra of strange particles. In few cases, the Monte Carlo
results are in qualitative agreement with the experimental data,
for which the values of $\chi^2$/ndof are quite large.

The $p_T$ spectra of $K^{+}+K^{-}$ (a),
$(K^{*0}+\overline{K}^{*0})/2$ (b) and $\phi$ (c) are displayed in
Figure~2 for Pb--Pb collisions at $\sqrt{s_{\rm NN}}=5.02$ TeV,
quoted from the ALICE Collaboration~\cite{46,47}. The
mid-(pseudo)rapidity is $|\eta|<0.2$ in panel (a), and $|y|<0.5$
in panels (b) and (c). Similarly to Figure 1, the data with
different centralities are represented by different symbols, and
the gray rectangles represent the uncertainty of the data. The
solid curves show the Monte Carlo results. For clarity, we use a
power index amplification from $10^1$ to $10^9$. The free
parameters $n$, $T_{0}$, $a_{0}$, and $\langle\beta_{t}\rangle$,
as well as $\chi^2$/ndof are listed in Table 2. One can see that
the Monte Carlo results show an approximate agreement with the
data in some cases, and a qualitative agreement with the data in
the other cases.

In addition to Pb--Pb collisions, the experimental $p_T$ spectra
in other collisions are also studied in order to better explore
the properties of strange particles. Figure 3 shows the $p_T$
spectra of $K^{+}+K^{-}$ (a), $(K^{*0}+\overline{K}^{*0})/2$ (b),
$K_S^0$ (c), $\phi$ (d), $\it\Lambda+\overline{\it\Lambda}$ (e),
$({\it\Xi^-}+\overline{{\it\Xi}}^+)/2$ (f), and
$({\it\Omega^-}+\overline{{\it\Omega}}^+)/2$ (g) in $p$--Pb
collisions at $\sqrt{s_{\rm NN}}=5.02$ TeV. The experimental data
in the panels are all quoted from the ALICE Collaboration when the
mid-rapidity is $0<y<0.5$ (a,c,e) and $-0.5<y<0$
(b,d,f,g)~\cite{48,49}. Different symbols indicate the different
centrality intervals. The Monte Carlo results are given by the
solid curves. The coefficients in parentheses are magnification
factors for better distinguishing the spectra in different
centrality intervals. The relevant parameters $n$, $T_{0}$,
$a_{0}$, and $\langle\beta_{t}\rangle$, as well as $\chi^{2}$/ndof
are listed in Table 3. One can see that the Monte Carlo results
are in good agreement with the data in most cases, and
approximately in agreement with the data in few cases.

Besides, we have also studied the $p_T$ spectra of $K^{+}+K^{-}$
(a) and $\phi$ (b) produced in Xe--Xe collisions at $\sqrt{s_{\rm
NN}}=5.44$ TeV in Figure 4. The symbols represent the data
measured by the ALICE Collaboration, and the mid-pseudorapidity is
$|\eta|<0.8$~\cite{50}. Similarly, the data with different
centralities are expressed as different symbols with gray
rectangles, and the solid curves show the Monte Carlo results.
Another, $10^n$ in legends are magnification factors for better
distinguishing the spectra. The relevant parameters $n$, $T_{0}$,
$a_{0}$, and $\langle\beta_{t}\rangle$, as well as $\chi^{2}$/ndof
are listed in Table 4. Once again, one can see that the Monte
Carlo results are in good agreement with the experimental data.

\begin{figure*}[htbp]
\begin{center}
\includegraphics[width=0.7\textwidth]{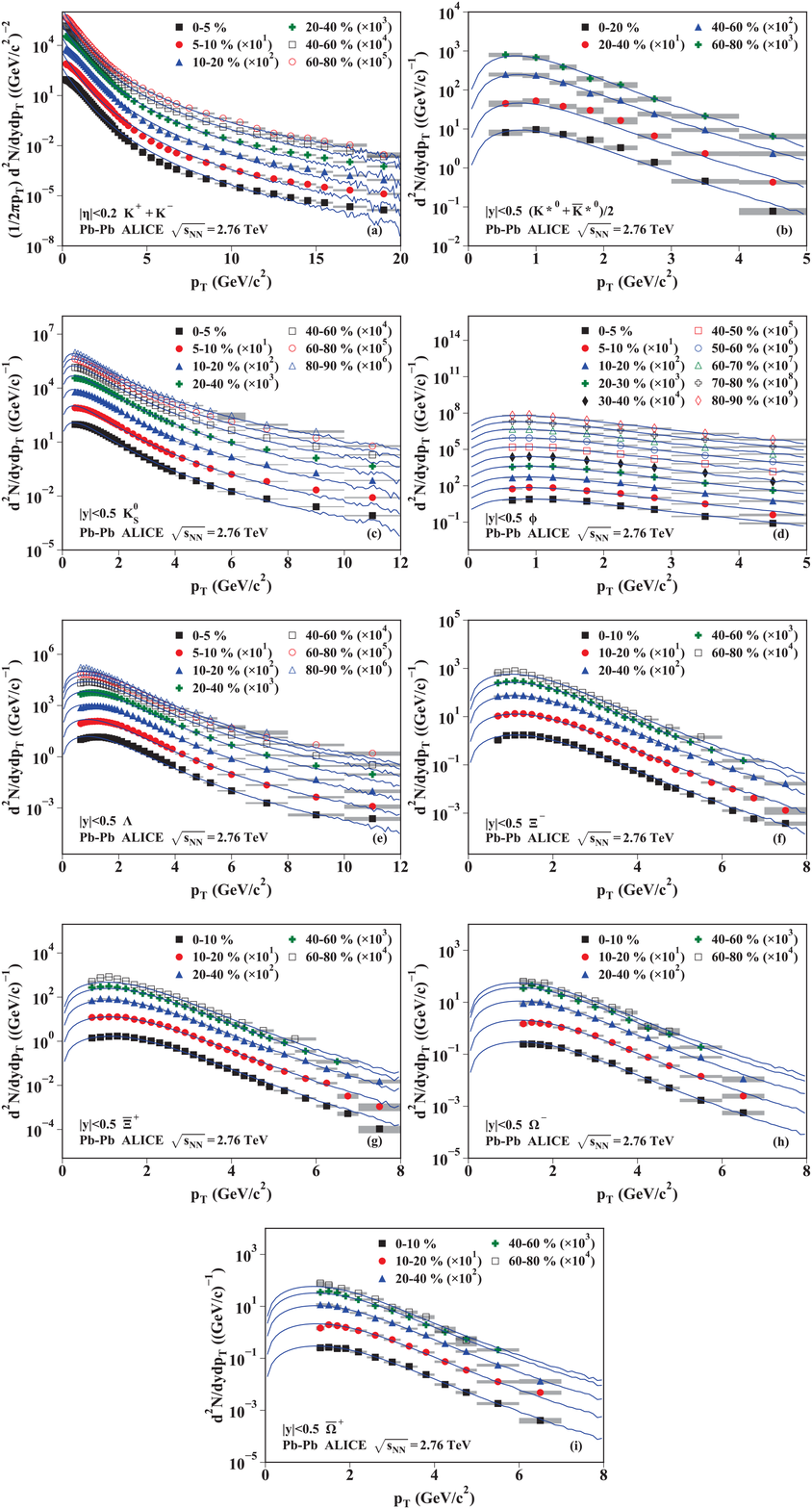}
\end{center}
\justifying\noindent {Figure 1. The $p_T$ spectra of $K^{+}+K^{-}$
(a), $(K^{*0}+\overline{K}^{*0})/2$ (b), $K_S^0$ (c), $\phi$ (d),
$\it\Lambda$ (e), $\it\Xi^-$ (f), $\overline{\it\Xi}^+$ (g),
$\it\Omega^-$ (h), and $\overline{\it\Omega}^+$ (i) in Pb--Pb
collisions with different centrality and mid-(pseudo)rapidity
intervals at $\sqrt{s_{\rm NN}}=2.76$ TeV. The symbols represent
the data measured by the ALICE Collaboration~\cite{43,44,45}. The
solid curves represent the Monte Carlo results, based on the
probability density function Eq. (2) and the $p_T$ expression in
Eq. (6).}
\end{figure*}

\begin{figure*}[htbp]
\begin{center}
\includegraphics[width=0.7\textwidth]{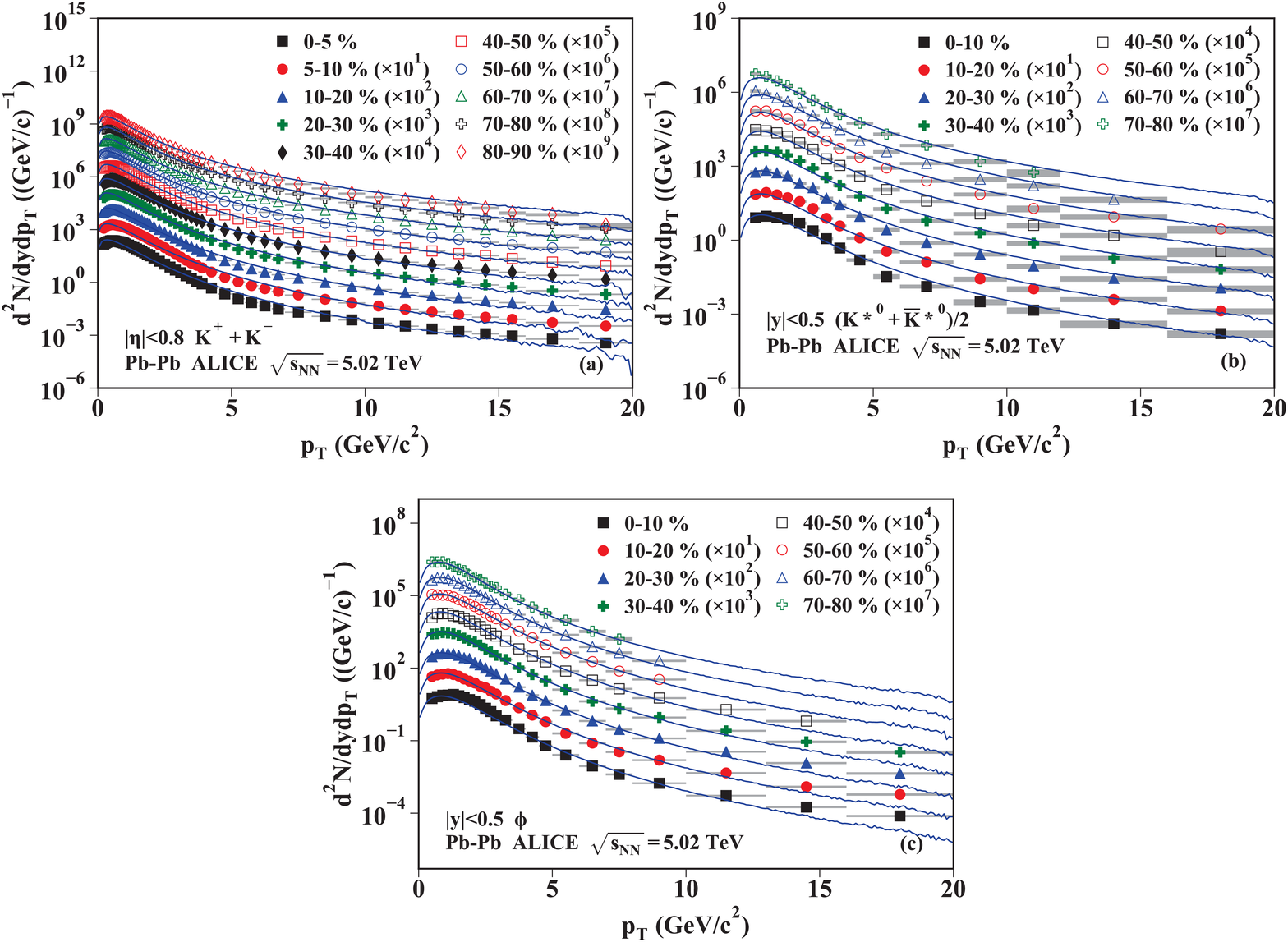}
\end{center}
\justifying\noindent {Figure 2. The $p_T$ spectra of $K^{+}+K^{-}$
(a), $(K^{*0}+\overline{K}^{*0})/2$ (b), and $\phi$ (c) in Pb--Pb
collisions at $\sqrt{s_{\rm NN}}=5.02$ TeV. The symbols represent
the data measured by the ALICE Collaboration~\cite{46,47}. The
solid curves represent the Monte Carlo results, based on the
probability density function Eq. (2) and the $p_T$ expression in
Eq. (6).}
\end{figure*}

\begin{figure*}[!htb]
\begin{center}
\includegraphics[width=0.7\textwidth]{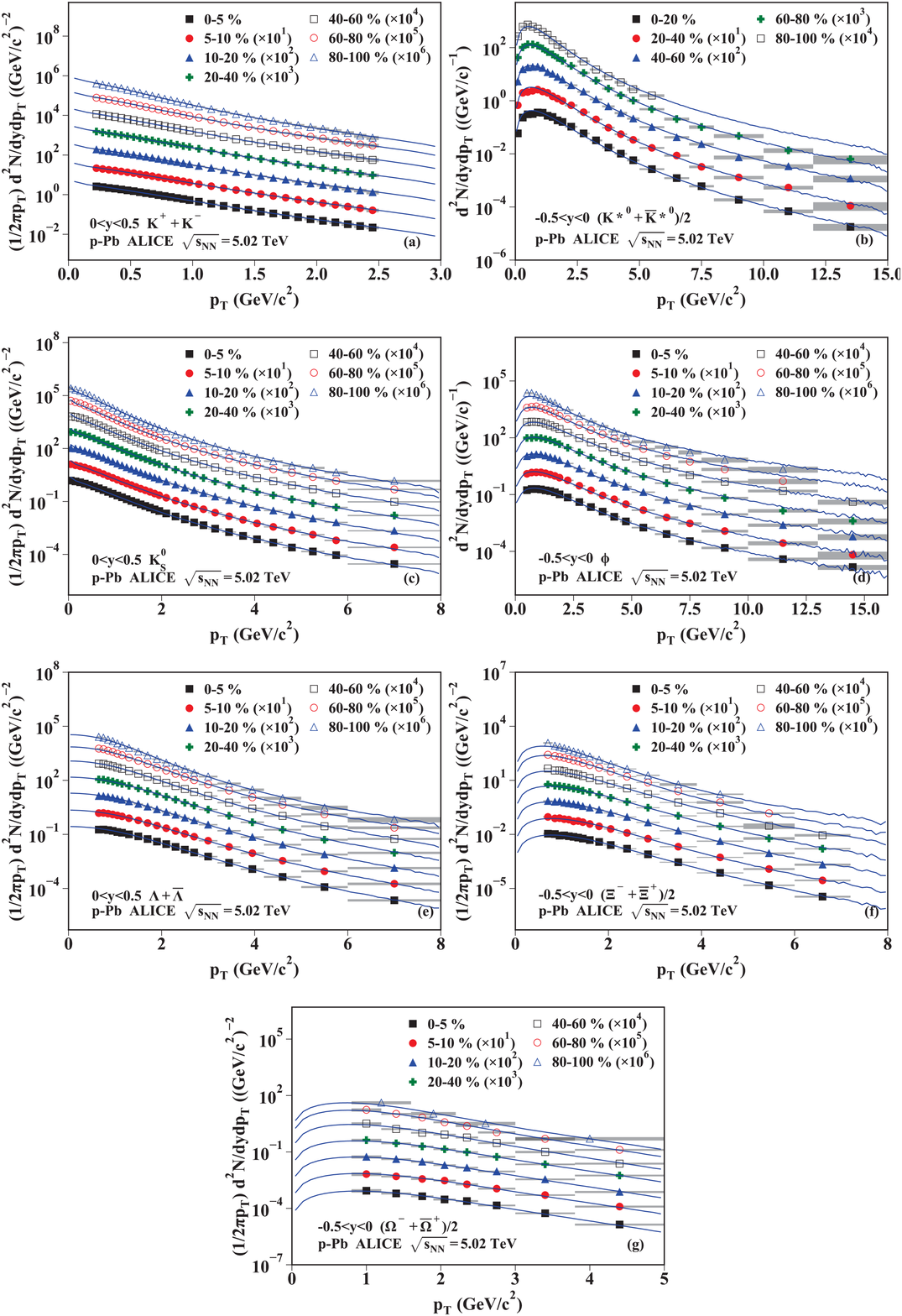}
\end{center}
\justifying\noindent {Figure 3. The $p_T$ spectra of $K^{+}+K^{-}$
(a), $(K^{*0}+\overline{K}^{*0})/2$ (b), $K_S^0$ (c), $\phi$ (d),
$\it\Lambda+\overline{\it\Lambda}$ (e),
$({\it\Xi^-}+\overline{{\it\Xi}}^+)/2$ (f), and
$({\it\Omega^-}+\overline{{\it\Omega}}^+)/2$ (g) in $p$--Pb
collisions at $\sqrt{s_{\rm NN}}=5.02$ TeV. The symbols represent
the data measured by the ALICE Collaboration~\cite{48,49}. The
solid curves represent the Monte Carlo results, based on the
probability density function Eq. (2) and the $p_T$ expression in
Eq. (6).}
\end{figure*}

\begin{figure*}[!htb]
\begin{center}
\includegraphics[width=0.7\textwidth]{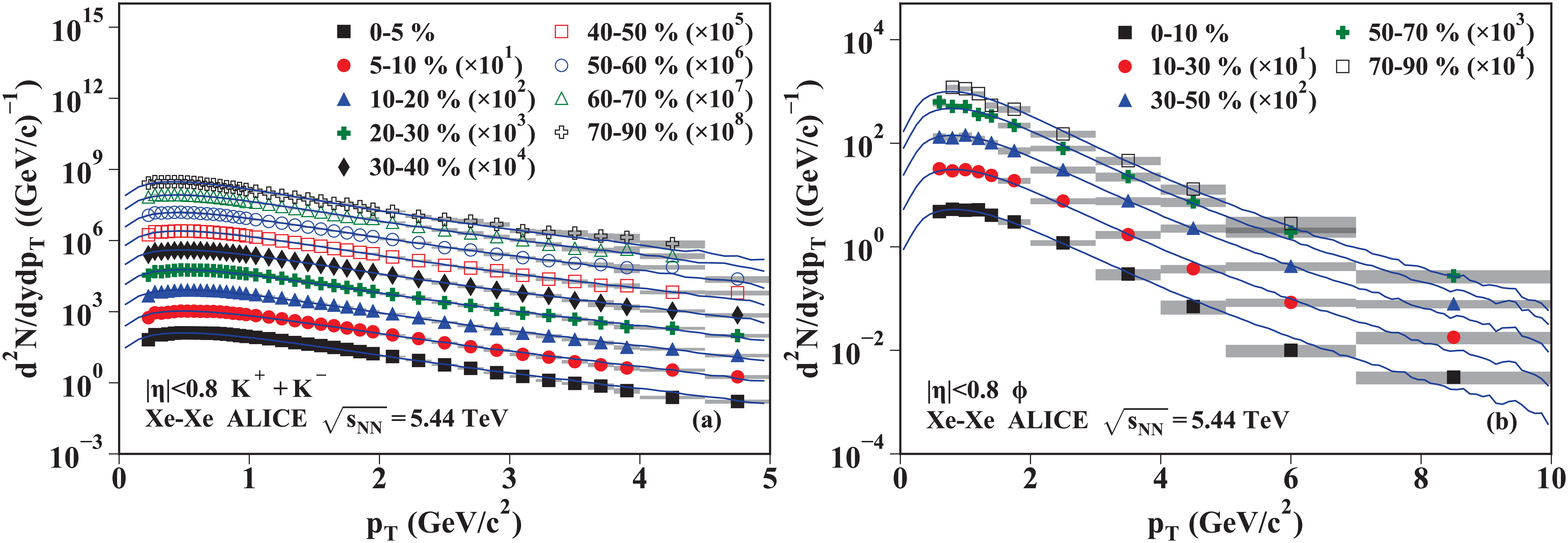}
\end{center}
\justifying\noindent {Figure 4. The $p_T$ spectra of $K^{+}+K^{-}$
(a) and $\phi$ (b) produced in Xe--Xe collisions at $\sqrt{s_{\rm
NN}}=5.44$ TeV. The symbols represent the data measured by the
ALICE Collaboration~\cite{50}. The solid curves represent the
Monte Carlo results, based on the probability density function Eq.
(2) and the $p_T$ expression in Eq. (6).}
\end{figure*}

\subsection{Tendencies of parameters}

\begin{table*} \vspace{0.25cm} \justifying\noindent {\small Table 1.
        Values of $n$, $T_{0}$, $a_{0}$, $\langle\beta_{t}\rangle$, $\chi^{2}$, and ndof corresponding to the solid curves in Figure
        1.} {\scriptsize
        \vspace{-0.35cm}

        \begin{center}
            \newcommand{\tabincell}[2]{\begin{tabular}{@{}#1@{}}#2\end{tabular}}
            \begin{tabular} {cccccccccccc}\\ \hline\hline
                Figure   & Particle &  $\sqrt{s_{\rm NN}}$ (TeV) & Selection & $n$ & $T_{0}$ (GeV) & $a_0$ & $\langle\beta_{t}\rangle$ ($c$) & $\chi^2$/ndof \\
                \hline
                Figure 1(a) & $K^{+}+K^{-}$  & $2.76 $     & 0--5\%  & $7.1\pm0.2$  & $0.221\pm0.003$ & $-0.022\pm0.004$ & $0.247\pm0.005$ & $1199/53$\\
                &                  & $|\eta|<0.2$& 5--10\%  & $7.0\pm0.2$  & $0.219\pm0.003$ & $-0.024\pm0.004$ & $0.246\pm0.005$ & $1021/53$\\
                &                  & Pb--Pb      & 10--20\% & $6.9\pm0.2$  & $0.216\pm0.003$ & $-0.026\pm0.004$ & $0.244\pm0.005$ & $887/53$\\
                &                  & $       $   & 20--40\% & $6.7\pm0.2$  & $0.212\pm0.003$ & $-0.029\pm0.004$ & $0.241\pm0.005$ & $793/53$\\
                &                  & $       $   & 40--60\% & $6.2\pm0.2$  & $0.209\pm0.003$ & $-0.032\pm0.004$ & $0.238\pm0.005$ & $323/53$\\
                &                  & $       $   & 60--80\% & $6.1\pm0.2$  & $0.208\pm0.003$ & $-0.035\pm0.004$ & $0.234\pm0.005$ & $115/53$\\
                \hline
                Figure 1(b) &$(K^{*0}+\overline{K}^{*0})/2$  & $2.76 $     & 0--20\%  & $7.7\pm0.2$  & $0.174\pm0.003$ & $1.042\pm0.010$ & $0.309\pm0.006$ & $21/3$\\
                &                                  & $|y|<0.5$   & 20--40\% & $7.4\pm0.2$  & $0.170\pm0.003$ & $1.037\pm0.010$ & $0.306\pm0.006$ & $22/3$\\
                &                                  & Pb--Pb      & 40--60\% & $7.2\pm0.2$  & $0.164\pm0.003$ & $1.032\pm0.010$ & $0.301\pm0.006$ & $3/3$\\
                &                                  & $       $   & 60--80\% & $6.9\pm0.2$  & $0.152\pm0.003$ & $1.027\pm0.010$ & $0.294\pm0.006$ & $3/3$\\
                \hline
                Figure 1(c) &$K^{0}_{S}$ & $2.76$   & 0--5\%  & $8.1\pm0.2$  & $0.179\pm0.003$ & $0.227\pm0.005$ & $0.310\pm0.006$ & $357/28$\\
                &           & $|y|<0.5$   & 5--10\%  & $8.0\pm0.2$  & $0.177\pm0.003$ & $0.226\pm0.005$ & $0.307\pm0.006$ & $337/28$\\
                &           & Pb--Pb      & 10--20\% & $7.8\pm0.2$  & $0.175\pm0.003$ & $0.224\pm0.005$ & $0.305\pm0.006$ & $348/28$\\
                &           & $       $   & 20--40\% & $7.5\pm0.2$  & $0.174\pm0.003$ & $0.220\pm0.005$ & $0.303\pm0.006$ & $318/28$\\
                &           & $       $   & 40--60\% & $7.1\pm0.2$  & $0.171\pm0.003$ & $0.218\pm0.005$ & $0.301\pm0.006$ & $221/28$\\
                &           & $       $   & 60--80\% & $6.7\pm0.2$  & $0.169\pm0.003$ & $0.214\pm0.005$ & $0.298\pm0.006$ & $151/28$\\
                &           & $       $   & 80--90\% & $6.4\pm0.2$  & $0.168\pm0.003$ & $0.210\pm0.005$ & $0.295\pm0.006$ & $46/28$\\
                \hline
                Figure 1(d) & $\phi$  & $2.76$      & 0--5\%  & $8.6\pm0.2$  & $0.183\pm0.003$ & $1.013\pm0.009$ & $0.311\pm0.006$ & $14/3$\\
                &           & $|y|<0.5$   & 5--10\%  & $8.5\pm0.2$  & $0.181\pm0.003$ & $1.012\pm0.009$ & $0.309\pm0.006$ & $31/3$\\
                &           & Pb--Pb      & 10--20\% & $8.3\pm0.2$  & $0.179\pm0.003$ & $1.011\pm0.009$ & $0.308\pm0.006$ & $20/3$\\
                &           & $       $   & 20--30\% & $8.2\pm0.2$  & $0.176\pm0.003$ & $1.009\pm0.009$ & $0.305\pm0.006$ & $7/3$\\
                &           & $       $   & 30--40\% & $7.9\pm0.2$  & $0.174\pm0.003$ & $1.006\pm0.009$ & $0.302\pm0.006$ & $8/3$\\
                &           & $       $   & 40--50\% & $7.7\pm0.2$  & $0.171\pm0.003$ & $0.999\pm0.009$ & $0.299\pm0.006$ & $5/3$\\
                &           & $       $   & 50--60\% & $7.4\pm0.2$  & $0.166\pm0.003$ & $0.996\pm0.009$ & $0.296\pm0.006$ & $3/3$\\
                &           & $       $   & 60--70\% & $7.1\pm0.2$  & $0.162\pm0.003$ & $0.991\pm0.009$ & $0.287\pm0.006$ & $1/3$\\
                &           & $       $   & 70--80\% & $6.9\pm0.2$  & $0.158\pm0.003$ & $0.987\pm0.009$ & $0.285\pm0.006$ & $4/3$\\
                &           & $       $   & 80--90\% & $6.6\pm0.2$  & $0.154\pm0.003$ & $0.982\pm0.009$ & $0.282\pm0.006$ & $2/3$\\
                \hline
                Figure 1(e) & $\it\Lambda$ & $2.76$      & 0--5\%  & $10.0\pm0.2$ & $0.137\pm0.002$ & $1.429\pm0.012$ & $0.218\pm0.004$ & $272/26$\\
                &             & $|y|<0.5$   & 5--10\%  & $9.8\pm0.2$  & $0.135\pm0.002$ & $1.426\pm0.012$ & $0.216\pm0.004$ & $211/26$\\
                &             & Pb--Pb      & 10--20\% & $9.7\pm0.2$  & $0.134\pm0.002$ & $1.420\pm0.012$ & $0.213\pm0.004$ & $197/26$\\
                &             & $       $   & 20--40\% & $9.5\pm0.2$  & $0.132\pm0.002$ & $1.419\pm0.012$ & $0.212\pm0.004$ & $127/26$\\
                &             & $       $   & 40--60\% & $9.2\pm0.2$  & $0.128\pm0.002$ & $1.416\pm0.012$ & $0.210\pm0.004$ & $32/26$\\
                &             & $       $   & 60--80\% & $9.0\pm0.2$  & $0.126\pm0.002$ & $1.413\pm0.012$ & $0.207\pm0.004$ & $38/26$\\
                &             & $       $   & 80--90\% & $8.8\pm0.2$  & $0.122\pm0.002$ & $1.410\pm0.012$ & $0.204\pm0.004$ & $60/24$\\
                \hline
                Figure 1(f) & $\it\Xi^{-}$ & $2.76 $      & 0--10\%  & $11.9\pm0.3$  & $0.137\pm0.002$ & $1.925\pm0.014$ & $0.238\pm0.005$ & $126/22$\\
                &             & $|y|<0.5$    & 10--20\% & $11.8\pm0.3$  & $0.135\pm0.002$ & $1.922\pm0.014$ & $0.236\pm0.005$ & $63/22$\\
                &             & Pb--Pb       & 20--40\% & $11.4\pm0.2$  & $0.132\pm0.002$ & $1.918\pm0.014$ & $0.233\pm0.005$ & $29/22$\\
                &             & $       $    & 40--60\% & $10.8\pm0.2$  & $0.126\pm0.002$ & $1.912\pm0.014$ & $0.230\pm0.005$ & $21/20$\\
                &             & $       $    & 60--80\% & $10.5\pm0.2$  & $0.120\pm0.002$ & $1.904\pm0.014$ & $0.227\pm0.005$ & $37/15$\\
                \hline
                Figure 1(g) & $\overline{\it\Xi}^{+}$ & $2.76$       & 0--10\%  & $12.1\pm0.3$  & $0.138\pm0.002$ & $1.924\pm0.014$ & $0.235\pm0.005$ & $122/22$\\
                &                        & $|y|<0.5$    & 10--20\% & $12.0\pm0.3$  & $0.136\pm0.002$ & $1.922\pm0.014$ & $0.233\pm0.005$ & $44/22$\\
                &                        & Pb--Pb       & 20--40\% & $11.8\pm0.3$  & $0.134\pm0.002$ & $1.918\pm0.014$ & $0.231\pm0.005$ & $33/22$\\
                &                        & $       $    & 40--60\% & $11.4\pm0.3$  & $0.130\pm0.002$ & $1.911\pm0.014$ & $0.227\pm0.005$ & $29/20$\\
                &                        & $       $    & 60--80\% & $11.1\pm0.2$  & $0.126\pm0.002$ & $1.910\pm0.014$ & $0.225\pm0.005$ & $72/15$\\
                \hline
                Figure 1(h) & $\it\Omega^{-}$ & $2.76 $     & 0--10\%  & $10.7\pm0.2$  & $0.133\pm0.002$ & $2.026\pm0.015$ & $0.248\pm0.005$ & $5/8$\\
                &                & $|y|<0.5$   & 10--20\% & $10.4\pm0.2$  & $0.131\pm0.002$ & $2.024\pm0.015$ & $0.247\pm0.005$ & $13/8$\\
                &                & Pb--Pb      & 20--40\% & $10.3\pm0.2$  & $0.129\pm0.002$ & $2.022\pm0.015$ & $0.245\pm0.005$ & $13/8$\\
                &                & $       $   & 40--60\% & $10.1\pm0.2$  & $0.126\pm0.002$ & $2.020\pm0.015$ & $0.242\pm0.005$ & $11/7$\\
                &                & $       $   & 60--80\% & $9.8\pm0.2$   & $0.124\pm0.002$ & $2.018\pm0.015$ & $0.237\pm0.005$ & $7/5$\\
                \hline
                Figure 1(i) &$\overline{\it\Omega}^{+}$ & $2.76$      & 0--10\%  & $10.9\pm0.2$  & $0.135\pm0.002$ & $2.022\pm0.015$ & $0.247\pm0.005$ & $7/8$\\
                &                          & $|y|<0.5$   & 10--20\% & $10.8\pm0.2$  & $0.134\pm0.002$ & $2.020\pm0.015$ & $0.245\pm0.005$ & $13/8$\\
                &                          & Pb--Pb      & 20--40\% & $10.6\pm0.2$  & $0.132\pm0.002$ & $2.017\pm0.015$ & $0.241\pm0.005$ & $6/8$\\
                &                          & $       $   & 40--60\% & $10.3\pm0.2$  & $0.130\pm0.002$ & $2.014\pm0.015$ & $0.239\pm0.005$ & $7/7$\\
                &                          & $       $   & 60--80\% & $9.8\pm0.2$   & $0.122\pm0.002$ & $2.004\pm0.015$ & $0.235\pm0.005$ & $7/5$\\
                \hline
            \end{tabular}%
    \end{center}}
\end{table*}

\begin{table*} \vspace{0.25cm} \justifying\noindent {\small Table 2.
        Values of $n$, $T_{0}$, $a_{0}$, $\langle\beta_{t}\rangle$, $\chi^{2}$, and ndof corresponding to the solid curves in Figure 2.
        \vspace{-0.35cm}

        \begin{center}
            \newcommand{\tabincell}[2]{\begin{tabular}{@{}#1@{}}#2\end{tabular}}
            \begin{tabular} {cccccccccccc}\\ \hline\hline
                Figure   & Particle &  $\sqrt{s_{\rm NN}}$ (TeV) & Selection & $n$ & $T_{0}$ (GeV) & $a_0$ & $\langle\beta_{t}\rangle$ ($c$) & $\chi^2$/ndof \\
                \hline
                Figure 2(a) & $K^{+}+K^{-}$  & $5.02 $   & 0--5\%  & $6.8\pm0.2$  & $0.228\pm0.003$ & $0.050\pm0.005$ & $0.251\pm0.005$ & $2716/48$\\
                &                & $|\eta|<0.8$& 5--10\%  & $6.7\pm0.2$  & $0.227\pm0.003$ & $0.048\pm0.005$ & $0.250\pm0.005$ & $2292/48$\\
                &                & Pb--Pb      & 10--20\% & $6.5\pm0.2$  & $0.225\pm0.003$ & $0.046\pm0.005$ & $0.248\pm0.005$ & $2239/48$\\
                &                & $       $   & 20--30\% & $6.3\pm0.2$  & $0.221\pm0.003$ & $0.042\pm0.005$ & $0.246\pm0.005$ & $1814/48$\\
                &                & $       $   & 30--40\% & $6.2\pm0.2$  & $0.218\pm0.003$ & $0.039\pm0.005$ & $0.244\pm0.005$ & $1468/48$\\
                &                & $       $   & 40--50\% & $5.9\pm0.2$  & $0.215\pm0.003$ & $0.035\pm0.005$ & $0.238\pm0.005$ & $996/48$\\
                &                & $       $   & 50--60\% & $5.7\pm0.2$  & $0.211\pm0.003$ & $0.032\pm0.005$ & $0.231\pm0.005$ & $621/48$\\
                &                & $       $   & 60--70\% & $5.5\pm0.2$  & $0.209\pm0.003$ & $0.029\pm0.005$ & $0.227\pm0.005$ & $267/48$\\
                &                & $       $   & 70--80\% & $5.3\pm0.2$  & $0.207\pm0.003$ & $0.027\pm0.005$ & $0.224\pm0.005$ & $219/48$\\
                &                & $       $   & 80--90\% & $5.1\pm0.2$  & $0.204\pm0.003$ & $0.023\pm0.005$ & $0.221\pm0.005$ & $487/48$\\
                \hline
                Figure 2(b) &$(K^{*0}+\overline{K}^{*0})/2$ & $5.02 $     & 0--10\%  & $7.0\pm0.2$  & $0.180\pm0.003$ & $1.028\pm0.010$ & $0.313\pm0.006$ & $103/10$\\
                &                                 & $|y|<0.5$   & 10--20\% & $6.8\pm0.2$  & $0.176\pm0.003$ & $1.034\pm0.010$ & $0.309\pm0.006$ & $81/10$\\
                &                                 & Pb--Pb      & 20--30\% & $6.6\pm0.2$  & $0.173\pm0.003$ & $1.018\pm0.010$ & $0.305\pm0.006$ & $95/10$\\
                &                                 & $       $   & 30--40\% & $6.5\pm0.2$  & $0.171\pm0.003$ & $1.015\pm0.010$ & $0.302\pm0.006$ & $72/10$\\
                &                                 & $       $   & 40--50\% & $6.4\pm0.2$  & $0.171\pm0.003$ & $1.011\pm0.010$ & $0.300\pm0.006$ & $88/10$\\
                &                                 & $       $   & 50--60\% & $6.3\pm0.2$  & $0.168\pm0.003$ & $1.008\pm0.010$ & $0.298\pm0.006$ & $66/10$\\
                &                                 & $       $   & 60--70\% & $6.2\pm0.2$  & $0.166\pm0.003$ & $1.005\pm0.010$ & $0.297\pm0.006$ & $74/9$\\
                &                                 & $       $   & 70--80\% & $6.2\pm0.2$  & $0.164\pm0.003$ & $1.000\pm0.010$ & $0.295\pm0.006$ & $32/8$\\
                \hline
                Figure 2(c) & $\phi$     & $5.02$   & 0--10\% & $8.1\pm0.2$  & $0.184\pm0.003$ & $1.017\pm0.010$ & $0.302\pm0.006$ & $323/19$\\
                &           & $|y|<0.5$   & 10--20\% & $8.0\pm0.2$  & $0.182\pm0.003$ & $1.014\pm0.010$ & $0.301\pm0.006$ & $137/19$\\
                &           & Pb--Pb      & 20--30\% & $7.9\pm0.2$  & $0.180\pm0.003$ & $1.009\pm0.010$ & $0.299\pm0.006$ & $144/19$\\
                &           & $       $   & 30--40\% & $7.7\pm0.2$  & $0.177\pm0.003$ & $1.006\pm0.010$ & $0.295\pm0.006$ & $87/19$\\
                &           & $       $   & 40--50\% & $7.6\pm0.2$  & $0.174\pm0.003$ & $1.002\pm0.009$ & $0.293\pm0.006$ & $77/18$\\
                &           & $       $   & 50--60\% & $7.4\pm0.2$  & $0.169\pm0.003$ & $0.996\pm0.009$ & $0.291\pm0.006$ & $12/16$\\
                &           & $       $   & 60--70\% & $7.2\pm0.2$  & $0.165\pm0.003$ & $0.991\pm0.009$ & $0.288\pm0.006$ & $16/16$\\
                &           & $       $   & 70--80\% & $6.9\pm0.2$  & $0.158\pm0.003$ & $0.987\pm0.009$ & $0.285\pm0.006$ & $16/15$\\
                \hline
            \end{tabular}%
    \end{center}}
\end{table*}

\begin{table*} \vspace{0.25cm} \justifying\noindent {\small Table 3.
        Values of $n$, $T_{0}$, $a_{0}$, $\langle\beta_{t}\rangle$, $\chi^{2}$, and ndof corresponding to the solid curves in Figure 3.
        \vspace{-0.35cm}

        \begin{center}
            \newcommand{\tabincell}[2]{\begin{tabular}{@{}#1@{}}#2\end{tabular}}
            \begin{tabular} {cccccccccccc}\\ \hline\hline
                Figure   & Particle &  $\sqrt{s_{\rm NN}}$ (TeV) & Selection & $n$ & $T_{0}$ (GeV) & $a_0$ & $\langle\beta_{t}\rangle$ ($c$) & $\chi^2$/ndof \\
                \hline
                Figure 3(a) & $K^{+}+K^{-}$ & $5.02 $    & 0--5\%  & $5.5\pm0.2$  & $0.223\pm0.003$ & $0.229\pm0.005$ & $0.257\pm0.005$ & $7/26$\\
                &                 & $0<y<0.5$  & 5--10\%  & $5.3\pm0.2$  & $0.215\pm0.003$ & $0.226\pm0.005$ & $0.252\pm0.005$ & $6/26$\\
                &                 & $p$--Pb    & 10--20\% & $5.1\pm0.2$  & $0.210\pm0.003$ & $0.221\pm0.005$ & $0.248\pm0.005$ & $4/26$\\
                &                 & $       $  & 20--40\% & $4.9\pm0.2$  & $0.199\pm0.003$ & $0.214\pm0.005$ & $0.241\pm0.005$ & $3/26$\\
                &                 & $       $  & 40--60\% & $4.7\pm0.2$  & $0.188\pm0.003$ & $0.205\pm0.005$ & $0.230\pm0.005$ & $2/26$\\
                &                 & $       $  & 60--80\% & $4.5\pm0.2$  & $0.167\pm0.003$ & $0.197\pm0.005$ & $0.221\pm0.005$ & $2/26$\\
                &                 & $       $  & 80--100\%& $4.3\pm0.2$  & $0.139\pm0.003$ & $0.183\pm0.005$ & $0.212\pm0.005$ & $5/26$\\
                \hline
                Figure 3(b) &$(K^{*0}+\overline{K}^{*0})/2$  & $5.02 $     & 0--20\%  & $6.8\pm0.2$  & $0.238\pm0.003$ & $0.550\pm0.007$ & $0.267\pm0.005$ & $72/17$\\
                &                                  & $-0.5<y<0$  & 20--40\% & $6.7\pm0.2$  & $0.231\pm0.003$ & $0.541\pm0.007$ & $0.258\pm0.005$ & $56/17$\\
                &                                  & $p$--Pb     & 40--60\% & $6.4\pm0.2$  & $0.225\pm0.003$ & $0.536\pm0.007$ & $0.254\pm0.005$ & $16/17$\\
                &                                  & $       $   & 60--80\% & $6.2\pm0.2$  & $0.211\pm0.003$ & $0.528\pm0.007$ & $0.249\pm0.005$ & $11/17$\\
                &                                  & $       $   & 80--100\%& $5.7\pm0.2$  & $0.183\pm0.003$ & $0.511\pm0.007$ & $0.233\pm0.005$ & $19/12$\\
                \hline
                Figure 3(c) &$K^{0}_{S}$ & $5.02$   & 0--5\%  & $6.6\pm0.2$  & $0.219\pm0.003$ & $0.202\pm0.005$ & $0.328\pm0.006$ & $78/29$\\
                &           & $0<y<0.5$   & 5--10\%  & $6.5\pm0.2$  & $0.216\pm0.003$ & $0.200\pm0.005$ & $0.325\pm0.006$ & $39/29$\\
                &           & $p$--Pb     & 10--20\% & $6.4\pm0.2$  & $0.214\pm0.003$ & $0.199\pm0.005$ & $0.323\pm0.006$ & $55/29$\\
                &           & $       $   & 20--40\% & $6.2\pm0.2$  & $0.209\pm0.003$ & $0.197\pm0.005$ & $0.321\pm0.006$ & $45/29$\\
                &           & $       $   & 40--60\% & $6.0\pm0.2$  & $0.204\pm0.003$ & $0.192\pm0.005$ & $0.317\pm0.006$ & $61/29$\\
                &           & $       $   & 60--80\% & $5.7\pm0.2$  & $0.189\pm0.003$ & $0.186\pm0.005$ & $0.312\pm0.006$ & $83/29$\\
                &           & $       $   & 80--100\%& $5.8\pm0.2$  & $0.174\pm0.003$ & $0.184\pm0.005$ & $0.304\pm0.006$ & $145/29$\\
                \hline
                Figure 3(d) & $\phi$  & $5.02$      & 0--5\%  & $6.7\pm0.2$  & $0.199\pm0.003$ & $0.812\pm0.009$ & $0.332\pm0.006$ & $18/15$\\
                &           & $-0.5<y<0$  & 5--10\%  & $6.6\pm0.2$  & $0.197\pm0.003$ & $0.810\pm0.009$ & $0.330\pm0.006$ & $26/15$\\
                &           & $p$--Pb     & 10--20\% & $6.4\pm0.2$  & $0.194\pm0.003$ & $0.806\pm0.009$ & $0.324\pm0.006$ & $29/15$\\
                &           & $       $   & 20--40\% & $6.3\pm0.2$  & $0.192\pm0.003$ & $0.803\pm0.009$ & $0.322\pm0.006$ & $25/15$\\
                &           & $       $   & 40--60\% & $6.1\pm0.2$  & $0.189\pm0.003$ & $0.799\pm0.008$ & $0.319\pm0.006$ & $20/15$\\
                &           & $       $   & 60--80\% & $5.5\pm0.2$  & $0.165\pm0.003$ & $0.783\pm0.008$ & $0.312\pm0.006$ & $38/14$\\
                &           & $       $   & 80--100\%& $5.1\pm0.2$  & $0.147\pm0.003$ & $0.775\pm0.008$ & $0.305\pm0.006$ & $63/14$\\
                \hline
                Figure 3(e) & $\it\Lambda+\overline{\it\Lambda}$ & $5.02$      & 0--5\%  & $7.7\pm0.2$  & $0.132\pm0.002$ & $1.426\pm0.012$ & $0.226\pm0.004$ & $6/15$\\
                &                                & $0<y<0.5$   & 5--10\%  & $7.6\pm0.2$  & $0.130\pm0.002$ & $1.424\pm0.012$ & $0.224\pm0.004$ & $4/15$\\
                &                                & $p$--Pb     & 10--20\% & $7.5\pm0.2$  & $0.127\pm0.002$ & $1.421\pm0.012$ & $0.221\pm0.004$ & $2/15$\\
                &                                & $       $   & 20--40\% & $7.3\pm0.2$  & $0.123\pm0.002$ & $1.417\pm0.012$ & $0.218\pm0.004$ & $4/15$\\
                &                                & $       $   & 40--60\% & $7.0\pm0.2$  & $0.112\pm0.002$ & $1.410\pm0.012$ & $0.211\pm0.004$ & $4/15$\\
                &                                & $       $   & 60--80\% & $6.7\pm0.2$  & $0.104\pm0.002$ & $1.407\pm0.012$ & $0.206\pm0.004$ & $22/15$\\
                &                                & $       $   & 80--100\%& $6.4\pm0.2$  & $0.089\pm0.002$ & $1.403\pm0.012$ & $0.202\pm0.004$ & $23/15$\\
                \hline
                Figure 3(f) & $({\it\Xi^{-}}+\overline{{\it\Xi}}^{+})/2$ & $5.02 $     & 0--5\%   & $9.0\pm0.2$  & $0.179\pm0.003$ & $0.703\pm0.008$ & $0.225\pm0.004$ & $15/12$\\
                &                                    & $-0.5<y<0$  & 5--10\%  & $8.9\pm0.2$  & $0.177\pm0.003$ & $0.701\pm0.008$ & $0.223\pm0.004$ & $14/12$\\
                &                                    & $p$--Pb     & 10--20\% & $8.4\pm0.2$  & $0.171\pm0.003$ & $0.692\pm0.008$ & $0.218\pm0.004$ & $9/12$\\
                &                                    & $       $   & 20--40\% & $8.0\pm0.2$  & $0.165\pm0.003$ & $0.688\pm0.008$ & $0.215\pm0.004$ & $18/12$\\
                &                                    & $       $   & 40--60\% & $7.2\pm0.2$  & $0.150\pm0.003$ & $0.678\pm0.008$ & $0.209\pm0.004$ & $40/12$\\
                &                                    & $       $   & 60--80\% & $7.0\pm0.2$  & $0.130\pm0.002$ & $0.669\pm0.008$ & $0.203\pm0.004$ & $19/11$\\
                &                                    & $       $   & 80--100\%& $6.4\pm0.2$  & $0.119\pm0.002$ & $0.662\pm0.008$ & $0.194\pm0.004$ & $25/10$\\
                \hline
                Figure 3(g) & $({\it\Omega^{-}}+\overline{{\it\Omega}}^{+})/2$ & $5.02 $     & 0--5\%   & $7.6\pm0.2$  & $0.174\pm0.003$ & $0.788\pm0.008$ & $0.240\pm0.005$ & $5/3$\\
                &                                          & $-0.5<y<0$  & 5--10\%  & $7.3\pm0.2$  & $0.173\pm0.003$ & $0.786\pm0.008$ & $0.238\pm0.005$ & $4/3$\\
                &                                          & $p$--Pb     & 10--20\% & $7.2\pm0.2$  & $0.170\pm0.003$ & $0.783\pm0.008$ & $0.235\pm0.005$ & $4/3$\\
                &                                          & $       $   & 20--40\% & $7.0\pm0.2$  & $0.167\pm0.003$ & $0.779\pm0.008$ & $0.232\pm0.004$ & $1/3$\\
                &                                          & $       $   & 40--60\% & $6.7\pm0.2$  & $0.150\pm0.003$ & $0.769\pm0.008$ & $0.223\pm0.004$ & $9/3$\\
                &                                          & $       $   & 60--80\% & $6.4\pm0.2$  & $0.141\pm0.002$ & $0.761\pm0.008$ & $0.214\pm0.004$ & $5/3$\\
                &                                          & $       $   & 80--100\%& $5.9\pm0.2$  & $0.132\pm0.002$ & $0.755\pm0.008$ & $0.206\pm0.004$ & $2/-$\\
                \hline
            \end{tabular}%
    \end{center}}
\end{table*}

\begin{table*} \vspace{0.25cm} \justifying\noindent {\small Table 4.
        Values of $n$, $T_{0}$, $a_{0}$, $\langle\beta_{t}\rangle$, $\chi^{2}$, and ndof corresponding to the solid curves in Figure 4.
        \vspace{-0.35cm}

        \begin{center}
            \newcommand{\tabincell}[2]{\begin{tabular}{@{}#1@{}}#2\end{tabular}}
            \begin{tabular} {cccccccccccc}\\ \hline\hline
                Figure   & Particle &  $\sqrt{s_{\rm NN}}$ (TeV) & Selection & $n$ & $T_{0}$ (GeV) & $a_0$ & $\langle\beta_{t}\rangle$ ($c$) & $\chi^2$/ndof \\
                \hline
                Figure 4(a) & $K^{+}+K^{-}$  & $5.44 $   & 0--5\%  & $8.3\pm0.2$  & $0.199\pm0.003$ & $0.399\pm0.006$ & $0.257\pm0.005$ & $73/33$\\
                &                & $|\eta|<0.8$& 5--10\%  & $8.1\pm0.2$  & $0.198\pm0.003$ & $0.396\pm0.006$ & $0.255\pm0.005$ & $58/33$\\
                &                & Xe--Xe      & 10--20\% & $7.9\pm0.2$  & $0.194\pm0.003$ & $0.395\pm0.006$ & $0.254\pm0.005$ & $55/33$\\
                &                & $       $   & 20--30\% & $7.7\pm0.2$  & $0.192\pm0.003$ & $0.392\pm0.006$ & $0.252\pm0.005$ & $37/33$\\
                &                & $       $   & 30--40\% & $7.5\pm0.2$  & $0.187\pm0.003$ & $0.390\pm0.006$ & $0.250\pm0.005$ & $20/33$\\
                &                & $       $   & 40--50\% & $7.3\pm0.2$  & $0.183\pm0.003$ & $0.386\pm0.006$ & $0.247\pm0.005$ & $16/33$\\
                &                & $       $   & 50--60\% & $7.2\pm0.2$  & $0.179\pm0.003$ & $0.384\pm0.006$ & $0.244\pm0.005$ & $14/33$\\
                &                & $       $   & 60--70\% & $7.1\pm0.2$  & $0.174\pm0.003$ & $0.381\pm0.006$ & $0.238\pm0.005$ & $16/32$\\
                &                & $       $   & 70--90\% & $6.9\pm0.2$  & $0.162\pm0.003$ & $0.368\pm0.006$ & $0.230\pm0.005$ & $27/32$\\
                \hline
                Figure 4(b) & $\phi$     & $5.44$   & 0--10\% & $9.0\pm0.2$  & $0.219\pm0.003$ & $0.810\pm0.009$ & $0.338\pm0.006$ & $13/6$\\
                &           & $|\eta|<0.8$& 10--30\% & $8.9\pm0.2$  & $0.216\pm0.003$ & $0.804\pm0.009$ & $0.334\pm0.006$ & $10/6$\\
                &           & Xe--Xe      & 30--50\% & $8.7\pm0.2$  & $0.213\pm0.003$ & $0.800\pm0.008$ & $0.332\pm0.006$ & $7/6$\\
                &           & $       $   & 50--70\% & $8.5\pm0.2$  & $0.209\pm0.003$ & $0.797\pm0.008$ & $0.327\pm0.006$ & $11/6$\\
                &           & $       $   & 70--90\% & $8.3\pm0.2$  & $0.203\pm0.003$ & $0.792\pm0.008$ & $0.317\pm0.006$ & $10/4$\\
                \hline
            \end{tabular}%
    \end{center}}
\end{table*}

To further understand the regularities shown by the $p_T$ spectra
of strange particles, we analyze the tendencies of various
parameters with changing centrality observed in the work. Figure 5
shows the dependence of entropy-related index $n$ on centrality
$C$ in Pb--Pb (a,b), $p$--Pb (c), and Xe--Xe (d) collisions at the
LHC energies. The symbols are quoted from Tables 1--4, where the
results from different particles are represented by different
symbols. As an entropy-related index, $n=1/(q-1)$ or $n=q'/(q'-1)$
describes the degree of non-equilibrium of the system, where $q$
($q'$) is the entropy index. Generally, a larger $n$ or a $q$
($q'$) closer to 1 corresponds to a higher degree of equilibrium.
The present work shows that $n$ is large enough or $q$ ($q'$) is
close to 1, and the system stays in an approximate equilibrium.
Meanwhile, the system in central collisions stays in higher degree
of equilibrium. In addition, as the rest mass of strange particle
increases, the value of $n$ increases. The multi-strange particles
correspond to larger $n$ than the single-strange particles. These
results imply that the system stays in larger degree of
equilibrium when it forms multi-strange particles.

Similar to Figure 5, Figure 6 shows the dependence of kinetic
freeze-out temperature $T_0$ on centrality $C$ in the mentioned
collisions at the LHC. One can see that $T_0$ decreases slightly
with the decrease of centrality from central to peripheral
collisions in most cases. In few cases, the decrease is
significant. In addition, $T_0$ extracted from the single-strange
particle spectra is larger than that from the multi-strange
particle spectra. This finding suggests that the single-strange
particles are formed earlier than the multi-strange particles,
though the latter may leave the system earlier than the former in
the hydrodynamic evolution due to different masses.

As shown in Figure 7, we present the changing law of the
correction index $a_{0}$ on centrality $C$ in the mentioned
collisions at the LHC. One can see that with the decrease of
centrality from central to peripheral collisions, $a_0$ is almost
invariant, though in few cases $a_0$ shows very slight decrease.
In most cases, the values of $a_0$ are far from 1, which means
that the introduction of $a_0$ is necessary. We have also compared
the present fits with $a_0\neq1$ with those by $a_0=1$. An obvious
difference appears in the low-$p_T$ region. This implies that the
fits by $a_0=1$ are not suitable. In fact, in the fits of the
present work, we use $a_0=1$ at the first, then we change $a_0$ to
fit the $p_T$ spectra if $a_0=1$ is not satisfactory.

The dependence of the average transverse flow velocity
$\langle\beta_{t}\rangle$ on centrality $C$ in the mentioned
collisions at the LHC is displayed in Figure 8. Similar
conclusions to the dependences of $n$ on $C$ and $T_0$ on $C$ can
be obtained. That is, $\langle\beta_{t}\rangle$ decreases slightly
with the decrease of centrality from central to peripheral
collisions in most cases. In few cases, the decrease is
significant. In addition, $\langle\beta_{t}\rangle$ extracted from
the single-strange particle spectra is larger than that from the
multi-strange particle spectra. This finding also suggests that
the single-strange particles are formed earlier than the
multi-strange particles.

In order to further study the final states of the strange
particles and the disorder degree of the system, we give the
results of pseudo-entropy $S'_{\rm hadron}=-\Sigma [f(p_T)/\Sigma
f(p_T)]\ln [f(p_T)/\Sigma f(p_T)]$~\cite{30}, which is based on
the probability density function $f(p_T)$ of $p_T$, inspired by
the entropy $S_{\rm hadron}=-\Sigma P(N)\ln P(N)$~\cite{51,52,53},
based on the probability density function $P(N)$ of multiplicity
$N$, where $p_T$ is in the units of GeV/$c$. As is done in our
recent work~\cite{29}, the width of $p_T$ bin is taken to be 0.1
GeV/$c$, though the width is changeable. The unit of $f(p_T)$ is
neglected due to the fact that $P(N)$ is dimensionless.

In Figure 9, we demonstrate the relation of the pseudoentropy
$S'_{\rm hadron}$ versus the centrality $C$. It is seen from the
results that $S'_{\rm hadron}$ decreases slightly with the
decrease of centrality from central to peripheral collisions in
most cases. Only in few cases, $S'_{\rm hadron}$ decreases
significantly with the decrease of $C$. The dependence of $S'_{\rm
hadron}$ on $C$ is similar to those of $n$, $T_0$ and
$\langle\beta_{t}\rangle$ on $C$. Generally, the multi-strange
particles show larger $S'_{\rm hadron}$ than the single-strange
particles, though both $S'_{\rm hadron}$ for multi- and
single-strange particles are negative. This also reflects the
mass-dependence of $S'_{\rm hadron}$, which is similar to those of
the other parameters.

Analyzing Figures 5--9, one can see that the correlations between
$n$ and $T_0$, $n$ and $\langle\beta_{t}\rangle$, $n$ and $S'_{\rm
hadron}$, $T_0$ and $\langle\beta_{t}\rangle$, $T_0$ and $S'_{\rm
hadron}$, and $\langle\beta_{t}\rangle$ and $S'_{\rm hadron}$ are
positive; while the correlations between $a_0$ and $n$, $a_0$ and
$T_0$, $a_0$ and $\langle\beta_{t}\rangle$, and $a_0$ and $S'_{\rm
hadron}$ are very small or negligible. These results are
understandable due to the fact that all $n$, $T_0$, and
$\langle\beta_{t}\rangle$ determine mainly, and $S'_{\rm hadron}$
is mostly defined by, the shape of the spectra in intermediate-
and high-$p_T$ regions; while $a_0$ determines mainly the spectra
in low- or even very low-$p_T$ region. Of course, because of the
requirement of normalization, all parameters affect the spectra in
the whole $p_T$ region, though the level of influence in different
$p_T$ regions are different.

\begin{figure*}[!htb]
\begin{center}
\includegraphics[width=0.7\textwidth]{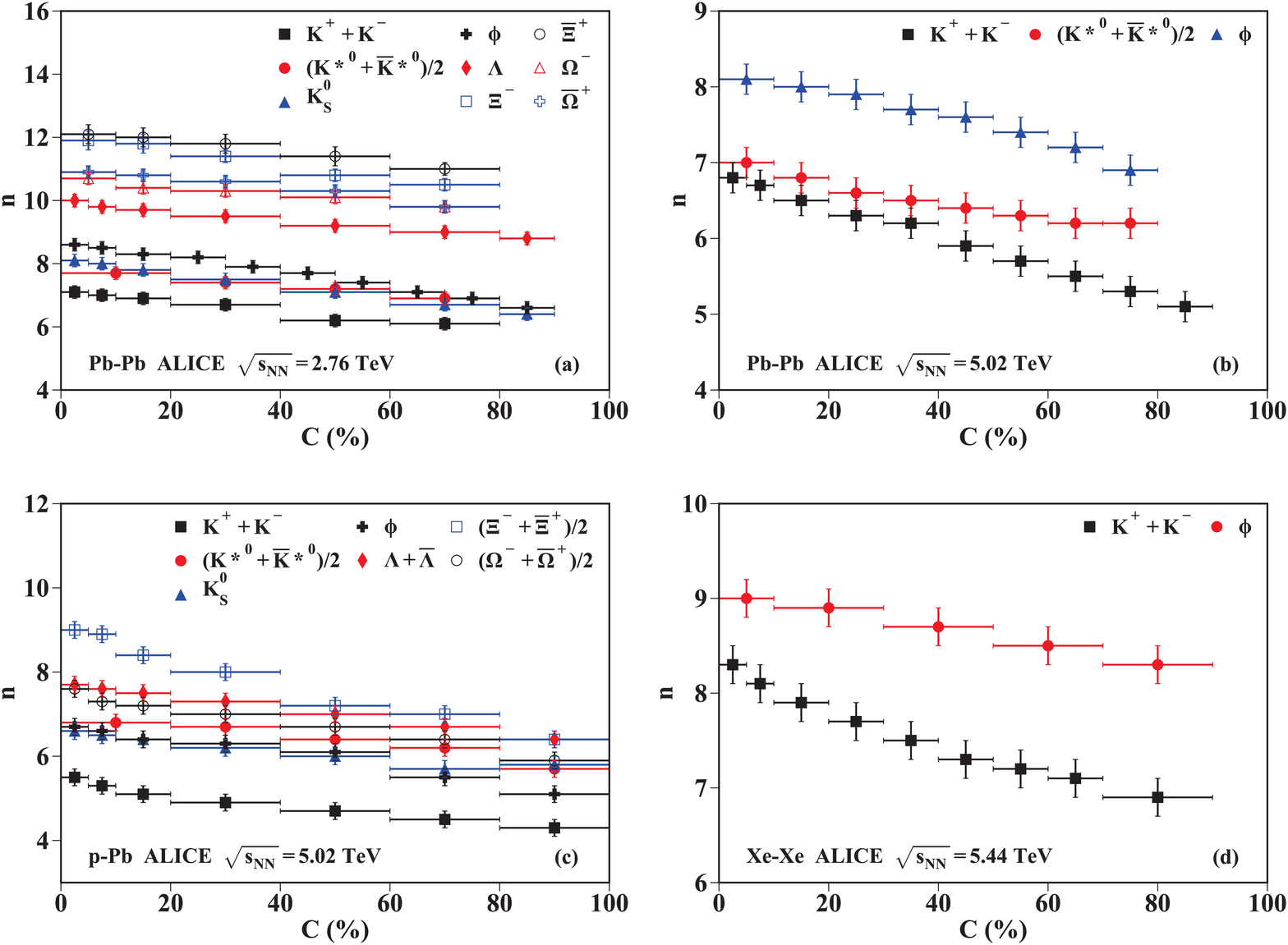}
\end{center}
\justifying\noindent {Figure 5. Dependence of entropy-related
index $n$ on centrality $C$ in Pb--Pb (a,b), $p$--Pb (c), and
Xe--Xe (d) collisions.}
\end{figure*}

\begin{figure*}[!htb]
\begin{center}
\includegraphics[width=0.7\textwidth]{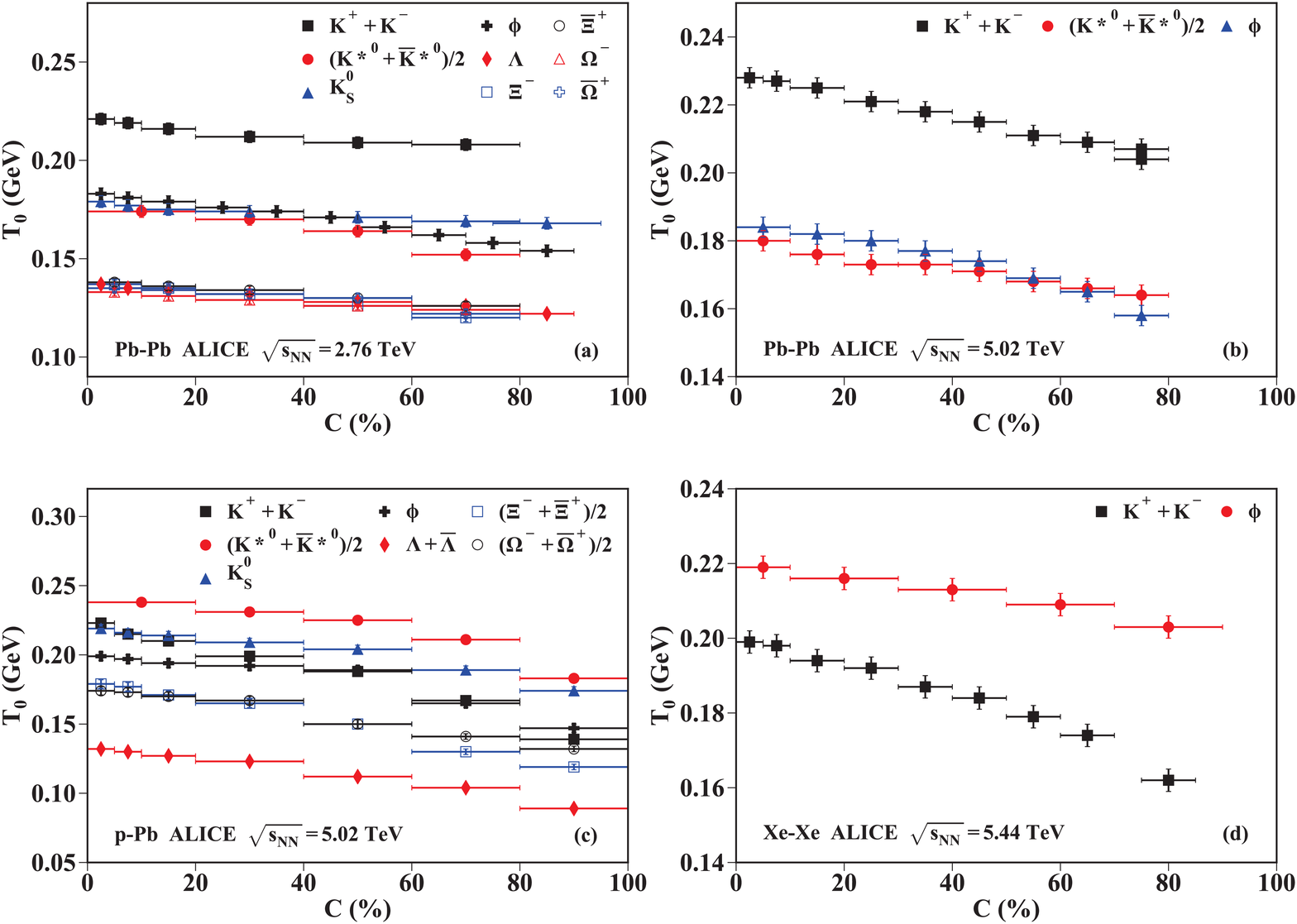}
\end{center}
\justifying\noindent {Figure 6. Dependence of kinetic freeze-out
temperature $T_0$ on centrality $C$ in Pb--Pb (a,b), $p$--Pb (c),
and Xe--Xe (d) collisions.}
\end{figure*}

\begin{figure*}[!htb]
\begin{center}
\includegraphics[width=0.7\textwidth]{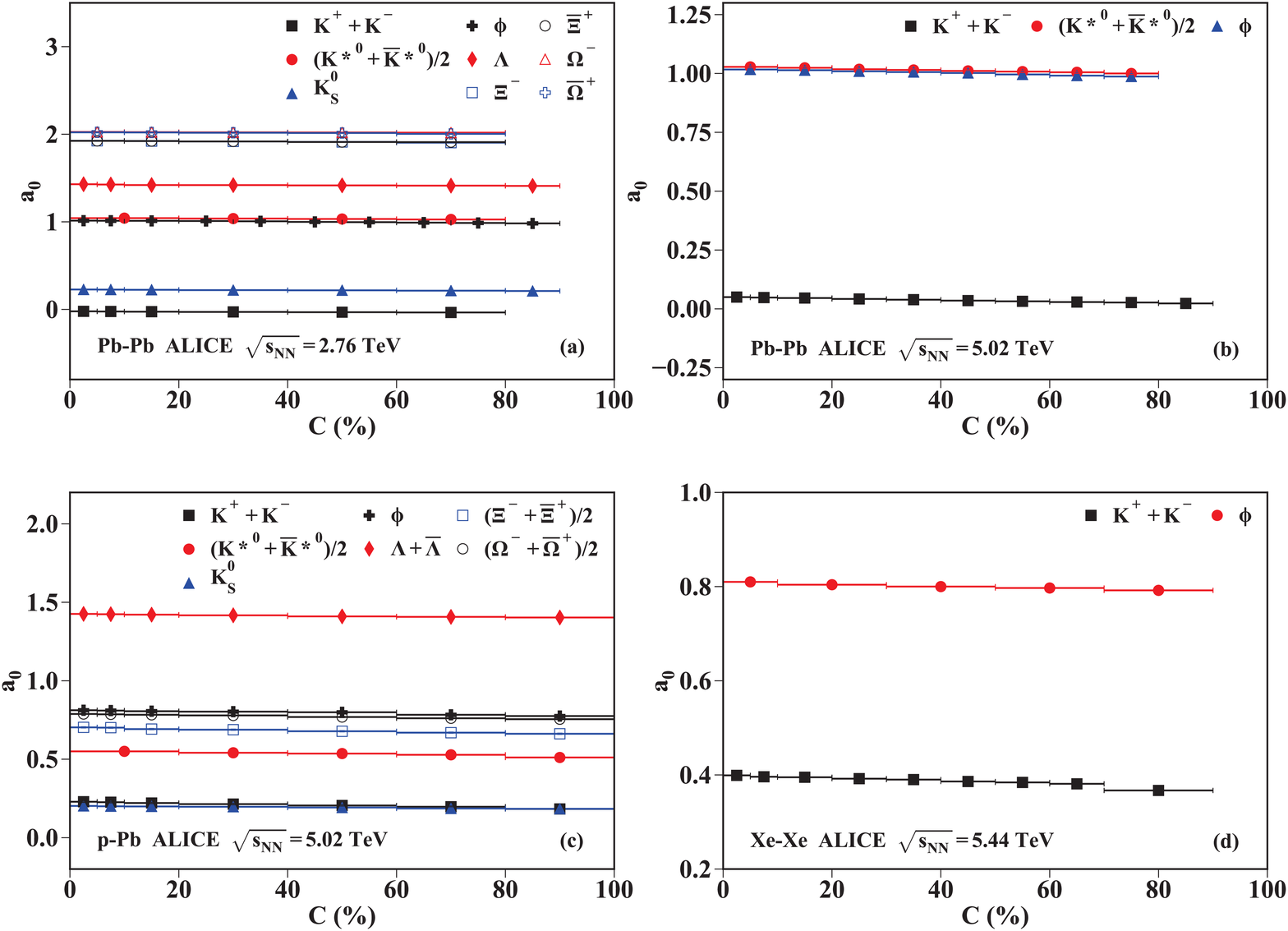}
\end{center}
\justifying\noindent {Figure 7. Dependence of correction index
$a_0$ on centrality $C$ in Pb--Pb (a,b), $p$--Pb (c), and Xe--Xe
(d) collisions.}
\end{figure*}

\begin{figure*}[!htb]
\begin{center}
\includegraphics[width=0.7\textwidth]{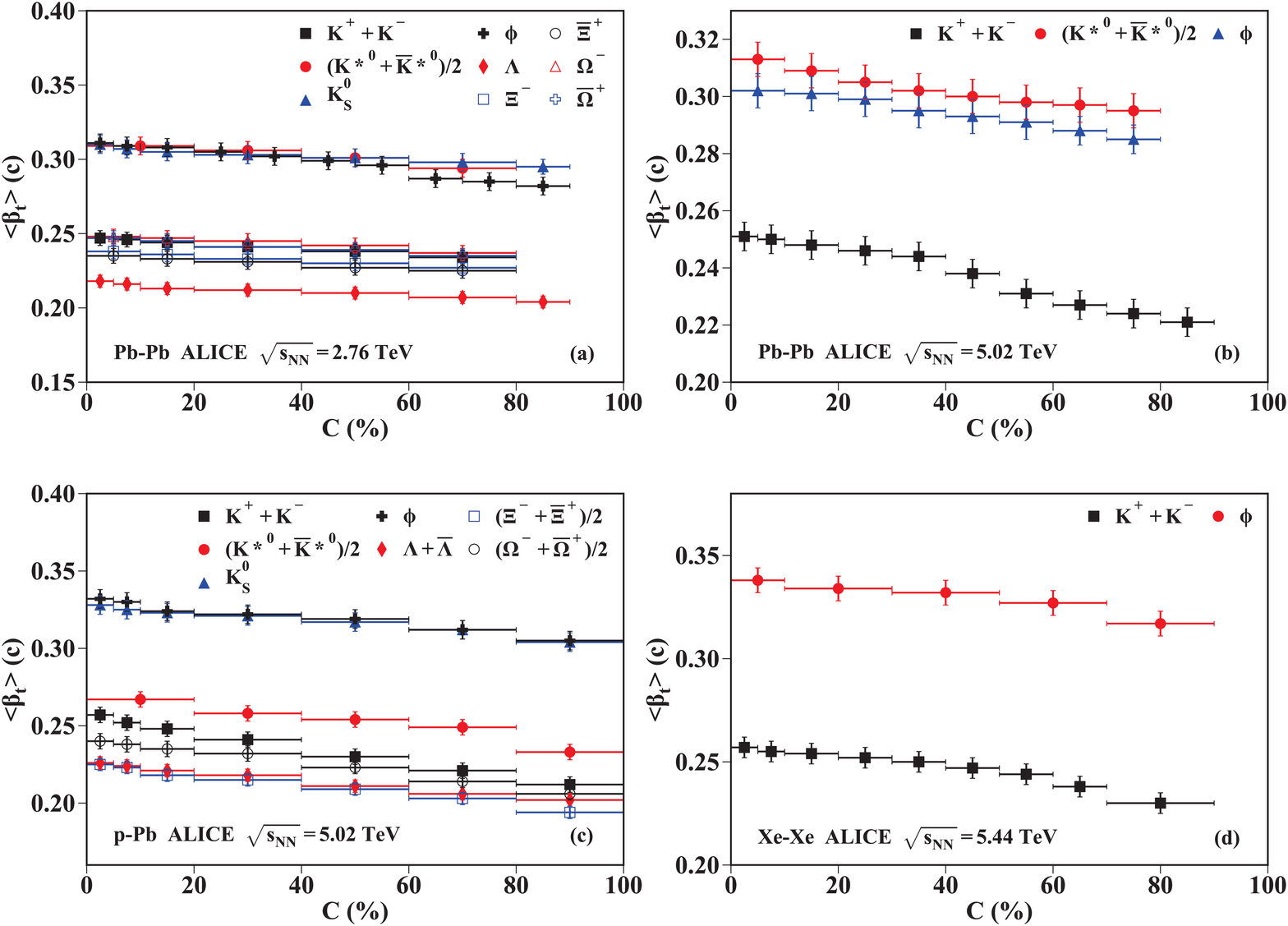}
\end{center}
\justifying\noindent {Figure 8. Dependence of average transverse
flow velocity $\langle\beta_{t}\rangle$ on centrality $C$ in
Pb--Pb (a,b), $p$--Pb (c), and Xe--Xe (d) collisions.}
\end{figure*}

\begin{figure*}[!htb]
\begin{center}
\includegraphics[width=0.7\textwidth]{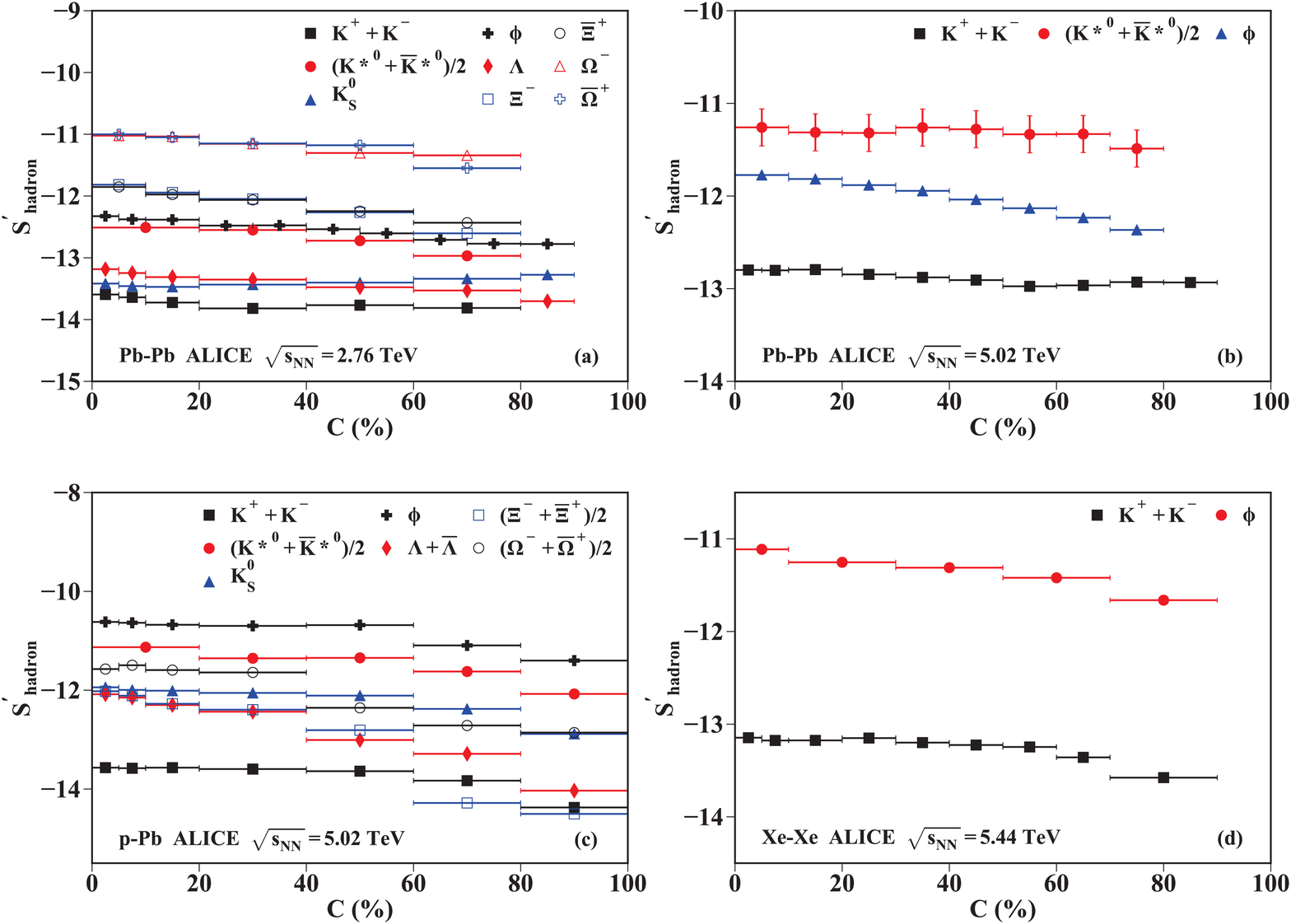}
\end{center}
\justifying\noindent {Figure 9. Dependence of pseudoentropy
$S'_{\rm hadron}$ on centrality $C$ in Pb--Pb (a,b), $p$--Pb (c),
and Xe--Xe (d) collisions.}
\end{figure*}

\subsection{Further discussion}

Although the extraction of $T_0$ and $\langle \beta_t\rangle$ in
present work is performed at the partonic-level, most extractions
in the literature have been done at the particle level, showing
inconsistent trends. For example, in terms of dependence of
parameters on centrality from central to peripheral collisions,
some works show an increase in $T_0$ and a decrease in
$\langle\beta_t\rangle$, and $\langle\beta_t\rangle\approx0$ in
small system or peripheral collisions~\cite{53a,53b,53c}. Other
work shows a decrease in both $T_0$ and $\langle\beta_t\rangle$,
and $\langle\beta_t\rangle$ is considerable in small system or
peripheral collisions~\cite{53d}. Generally, the relative
difference between two $T_0$ in central and peripheral collisions
is small ($\sim10\%$).

We observe a slightly higher $T_0$ in central collisions at both
the partonic- and particle level. If so, the centrality dependence
of $T_0$ is similar to those of the chemical freeze-out
temperature and effective temperature. However, it is still an
open question which $T_0$ is larger. In any case, the explanation
is understandable. A higher $T_0$ in central collisions means a
higher excitation, and a lower $T_0$ in central collisions means a
longer lifetime, of the hot and dense matter. Although we may
extract the thermal parameters at the particle level, we aim to
perform it at the partonic-level. The reason is that the
similarity, commonality, and universality available in high energy
collisions~\cite{53e4,53e5,53e6,53e7,53e8,53e9,53e10,53e11} should
be related to partons, which are a deeper level in the structure
of matter compared to the particle level.

The present work shows that central collisions correspond to a
higher excitation degree and a larger blast due to the more energy
being deposited. Meanwhile, central collisions correspond to
higher degree of equilibrium due to many particles being produced.
Compared with the parameters at the particle level, the parameters
at the partonic-level are extracted at earlier time moment and
correspond to larger $T_0$ and $\langle \beta_t\rangle$. However,
the system at the particle level is at higher degree of
equilibrium due to the fact that longer time is taken to approach
the chemical and kinetic freeze-out. However, because the time
span from parton phase to particle phase is very small, the
difference between two sets of parameters at the partonic- and
particle level is very small.

In some cases, the fitting quality is not so good due to large
$\chi^2$ values. This is because only experimental uncertainty is
considered. In fact, we have used the Monte Carlo method, which
causes additional uncertainty. If we take the uncertainty induced
by the Monte Carlo method to be approximately the same as the
experimental uncertainty, $\chi^2$ values will be reduced by
$1/\sqrt{2}\approx70.7$\%. In the case of large $\chi^2$ values,
we may regard the fits as the qualitative ones. In most cases, the
fits are good or approximate, and in few cases the fits are
qualitative.

Before summarizing and concluding, we would like to conduct some
further discussions on the Monte Carlo results. As we know, most
of baryons consist of up and down quarks, and the number of
strange quarks is very small. Only when there is a formation of
QGP, the system is likely to produce a lot of strange quarks and
strange anti-quarks, and then, strange quarks and strange
anti-quarks are able to combine with other neighbouring partons to
form strange particles. Thus, the abundant yield of the
final-state strange particles is an important signal of the
existence of QGP matter~\cite{54,55,56,57,58,59}. Because of the
formation of QGP with abundance of strange particles, the
implementation of this work has become possible due to sufficient
statistics.

In the above discussions, we have used different sets of the
parameters to fit the $p_T$ spectra of different strange
particles. This means that we have used the multi-scenario of
kinetic freeze-out. In some cases, the two-scenario of kinetic
freeze-out is also applicable, if we consider the single-strange
and multi-strange particles, respectively. In our opinion, the
single-scenario is a rough description, the two-scenario is a
slight refined description, and the multiple-scenario is a more
refined description, of the process of kinetic freeze-out. This
situation is analogous to the atomic spectra and their fine
structures.

\section{Summary and Conclusions}

The transverse momentum spectra of strange particles, including
$K^{\pm}$, $\phi$, $\it\Omega$, $\it\Xi$, and so on, in high
energy collisions are analyzed by considering the contributions of
constituent quarks. Each constituent quark contributes to the
transverse momentum obeying the modified Tsallis--Pareto-type
function with random azimuthal angle. The transverse momentum of
strange particle is the vector superposition of the transverse
momenta of the two or three constituent quarks. The results
calculated by the Monte Carlo method are in good agreement with
the experimental data in most cases in Pb--Pb, $p$--Pb, and Xe--Xe
collisions at a few TeV, measured by the ALICE Collaboration at
the LHC. In few cases, the agreement is qualitative due to quite
large $\chi^2$ values.

With the decrease of centrality from central to peripheral
collisions, the free parameters $n$, $T_0$, and
$\langle\beta_{t}\rangle$, as well as the derived parameter
$S'_{\rm hadron}$ decrease slightly in most cases, and decrease
significantly in few cases. Meanwhile, the free parameter $a_0$ is
almost invariant in most cases, and decreases slightly in few
cases. These results imply that central collisions stay in the
state with larger degree of equilibrium, higher excitation, and
larger blast than peripheral collisions, though both the central
and peripheral collisions stay in approximate equilibrium when the
system produces a lot of strange particles.

The correlations between $n$ and $T_0$, $n$ and
$\langle\beta_{t}\rangle$, $n$ and $S'_{\rm hadron}$, $T_0$ and
$\langle\beta_{t}\rangle$, $T_0$ and $S'_{\rm hadron}$, and
$\langle\beta_{t}\rangle$ and $S'_{\rm hadron}$ are positive,
because all $n$, $T_0$, and $\langle\beta_{t}\rangle$ determine
mainly, and $S'_{\rm hadron}$ is mostly defined by the shape of
the spectra in intermediate- and high-$p_T$ regions. The
correlations between $a_0$ and $n$, $a_0$ and $T_0$, $a_0$ and
$\langle\beta_{t}\rangle$, and $a_0$ and $S'_{\rm hadron}$ are
very small or negligible, because $a_0$ determines mainly the
spectra in (very) low-$p_T$ region.

We have used the multi-scenario of kinetic freeze-out, though in
some cases the two-scenario of kinetic freeze-out may also be
applicable if we consider a part of $p_T$ region. Meanwhile, if we
consider the single-strange and multi-strange particles
respectively, the single-strange particles are shown to form
earlier than the multi-strange ones. We may regard the relations
of various scenarios as follows: the single-scenario is a rough
description, the two-scenario is a slightly refined description,
and the multiple-scenario is a more refined description of the
process of kinetic freeze-out.
\\
\\
{\bf Data Availability}

The data used to support the findings of this study are included
within the article and are cited at relevant places within the
text as references.
\\
\\
{\bf Ethical Approval}

The authors declare that they are in compliance with ethical
standards regarding the content of this paper.
\\
\\
{\bf Disclosure}

The funding agencies have no role in the design of the study; in
the collection, analysis, or interpretation of the data; in the
writing of the manuscript; or in the decision to publish the
results.
\\
\\
{\bf Conflicts of Interest}

The authors declare that there are no conflicts of interest
regarding the publication of this paper.
\\
\\
{\bf Acknowledgments}

The work of X.-H.Z. was supported by the Innovative Foundation for
Graduate Education in Shanxi University. The work of Shanxi Group
was supported by the National Natural Science Foundation of China
under Grant Nos. 12147215, 12047571, and 11575103, the Shanxi
Provincial Natural Science Foundation under Grant Nos.
202103021224036 and 201901D111043, the Scientific and
Technological Innovation Programs of Higher Education Institutions
in Shanxi (STIP) under Grant No. 201802017, and the Fund for
Shanxi ``1331 Project" Key Subjects Construction. The work of
Kh.K.O. was supported by the Ministry of Innovative Development of
the Republic of Uzbekistan within the fundamental project No.
F3-20200929146 on analysis of open data on heavy-ion collisions at
RHIC and LHC. The work of A.D. was partially supported by the
Conselho Nacional de Desenvolvimento Cient{\'i}fico e
Tecnol{\'o}gico (CNPq-Brazil), by Project INCT-FNA Proc. No. 464
898/2014-5, and by FAPESP under grant 2016/17612-7.
\\
\\

\end{document}